\documentclass[twocolumn, prb, showpacs, english, preprintnumbers, amsmath, amssymb, superscriptaddress, aps]{revtex4-1}
\usepackage{latexsym}
\usepackage{epsfig,psfrag}
\usepackage{dcolumn}
\usepackage{bm}
\usepackage{graphicx}
\usepackage{color}
\usepackage{esint}
\usepackage{babel}
\usepackage{amsfonts}
\usepackage{enumerate}
\usepackage{mathbbol}

\begin{document}
\title{Proximity-induced superconductivity in Landau-quantized graphene monolayers} 

\author{Laura Cohnitz}
\affiliation{Institut f\"ur Theoretische Physik, 
Heinrich-Heine-Universit\"at, D-40225 D\"usseldorf, Germany}

\author{Alessandro De Martino}
\affiliation{Department of Mathematics, City, University of London, 
London EC1V 0HB, United Kingdom}

\author{Wolfgang H\"ausler}
\affiliation{Institut f\"ur Physik, 
Universit\"at Augsburg, D-86135 Augsburg, Germany}
\affiliation{I. Institut f\"ur Theoretische Physik, 
Universit\"at Hamburg, D-20355 Hamburg, Germany}

\author{Reinhold Egger}
\affiliation{Institut f\"ur Theoretische Physik, 
Heinrich-Heine-Universit\"at, D-40225 D\"usseldorf, Germany}

\date{\today}

\begin{abstract}
We consider massless Dirac fermions in a graphene monolayer 
in the ballistic limit, subject to both a perpendicular 
magnetic field $B$ and 
a proximity-induced pairing gap $\Delta$. When the chemical potential is at the 
Dirac point, our exact solution of the Bogoliubov-de Gennes equation 
yields $\Delta$-independent relativistic Landau levels.  Since eigenstates
depend on $\Delta$, many observables nevertheless are sensitive to pairing, e.g.,
the local density of states or the edge state spectrum.  By solving the problem 
with an additional in-plane electric field, we also discuss how 
snake states are influenced by a pairing gap. 
\end{abstract}

\maketitle

\begin{figure}
\centering
\includegraphics[width=0.45\textwidth]{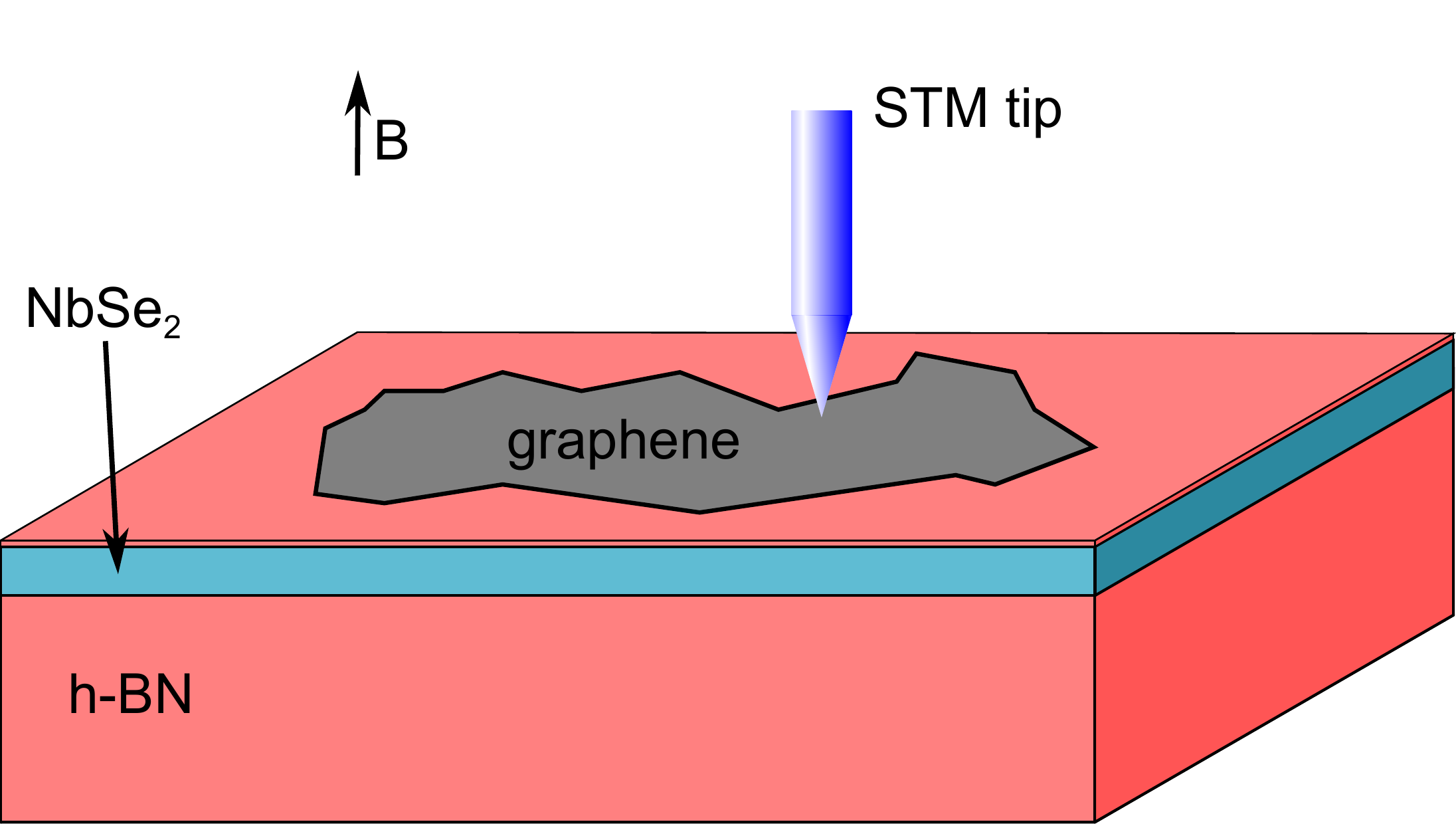} 
\caption{\label{fig1} Sketch of a vertical hybrid structure in a perpendicular
magnetic field $B$, where the graphene flake is deposited on a superconducting film 
(e.g., a few monolayers of NbSe$_2$) supported by an h-BN substrate.  
Inserting an h-BN monolayer between the superconductor and the graphene sample  
allows to gate the device (gates not shown).   
The graphene layer may be probed by an STM tip as indicated. 
Alternatively, the stack could be closed by a top h-BN monolayer.  }
\end{figure}

\emph{Introduction.---}It is well known that at energies close to the neutrality point, the
 electronic properties of graphene monolayers are accurately 
 described in terms of two-dimensional (2D) massless Dirac fermions 
\cite{Geim2004,Geim2005,Beenakker2008,CastroNeto2009,Goerbig2011,
Andrei2012,Miransky2015}. Recent advances in fabrication and 
preparation technology \cite{Andrei2012,Dean2010} allow
experimentalists to routinely reach the ballistic (disorder-free) transport regime.
Our theoretical work reported below is largely motivated by spectacular recent
 progress on Josephson transport in 
 ballistic graphene flakes contacted by conventional superconductors  
\cite{Calado2015,Lee2015,Allen2016,BenShalom2016,Efetov2016,Borzenets2016,
Amet2016,Zhu2017,Nanda2017,Lee2017,Bretheau2017}, demonstrating 
in particular that proximity-induced superconductivity can coexist with rather
high (Landau-quantizing) magnetic fields 
\cite{BenShalom2016,Amet2016,Lee2017}.
 This raises the question of  how a proximity-induced \textit{bulk}\ pairing gap 
 will affect the electronic properties of graphene in an orbital magnetic field. 
 In contrast to lateral graphene-superconductor interfaces, where theory is well developed~\cite{Beenakker2008,Beenakker2006,Titov2006,Ossipov2007}, 
we therefore investigate vertical hybrid structures as shown schematically
in Fig.~\ref{fig1}.  Superconductivity can be proximity-induced in the graphene 
sample from a 2D van der Waals superconductor \cite{Christian}, e.g., using a NbSe$_2$ film supported on a standard hexagonal boron nitride (h-BN) substrate \cite{Dean2010}. 
NbSe$_2$ is a good superconductor with high critical field ($B_{c2}\approx 5$~T at $T=1$~K), remains superconducting down to a few monolayers, and exhibits 
high-quality interfaces with graphene \cite{Efetov2016}.
For gating the device, another h-BN monolayer may be inserted as indicated 
in Fig.~\ref{fig1}, at the expense of reducing the proximity gap.  
The proximitized graphene flake can be probed by a 
scanning tunneling microscope (STM), e.g., using a graphite finger tip for 
ultra-high energy resolution \cite{Bretheau2017}.

Before turning to derivations, we briefly summarize our main results which can be 
tested by established STM techniques \cite{Zhao2015}, transport experiments, 
and/or local manipulation of defect charges in the substrate \cite{Lee2016}:
(i) By means of an exact solution of the  
Bogoliubov-de Gennes (BdG) equation, we show that at the Dirac point, i.e., 
for chemical potential $\mu=0$, the 
energy spectrum of a proximitized graphene layer in a homogeneous 
magnetic field $B$ is independent of the proximity gap $\Delta$.  
The BdG spectrum thus reduces to the familiar
relativistic Landau level spectrum \cite{CastroNeto2009}, in marked difference to
 the time-reversal-symmetric case with a strain-induced 
pseudo-magnetic field where the spectrum depends on $\Delta$ in a conventional 
manner \cite{Uchoa2013,Roy2014,LeeNandi2017}.  
(ii)  Even though the energy spectrum is independent of $\Delta$ at the Dirac point, 
 the corresponding eigenstates are sensitive to the pairing gap. 
Clear experimental signatures of proximity-induced superconductivity in Landau-quantized graphene are predicted for the energy-resolved local density of states 
(DOS) as well as for the edge states present near the sample boundaries.
 Away from the Dirac point, also the spectrum itself depends on $\Delta$.
(iii) Chiral snake-like states are expected in graphene for $\Delta=0$ in the presence of 
a weak electric field $\cal E$ perpendicular to $B$ \cite{Lukose2007,Liu2015,Cohnitz2016},
see Refs.~\cite{Tatch2015,Rickshaus2015} for recent experimental reports.
We  solve the corresponding BdG equation for arbitrary $\Delta$ 
through a Lorentz transformation of our solution for case (i), 
and thereby discuss how snake states are affected by a pairing gap.   

\emph{Model.---}We start from the
BdG equation, $H \Psi= E\Psi$, for proximitized graphene samples as in Fig.~\ref{fig1}.
The BdG Hamiltonian is represented by the matrix \cite{Beenakker2008,Beenakker2006}, 
\begin{equation}\label{dbdg}
H=\left(\begin{array}{cc}  v_F 
\left (\bm{\hat p}+\frac{e}{c}\bm{A} \right)\cdot \bm{\sigma} + V & 
\Delta \\ \Delta^\ast & 
-v_F \left ( \bm{\hat p}-\frac{e}{c}\bm{A} \right)\cdot \bm{\sigma} - V
\end{array}\right),
\end{equation}
with canonical momentum $\bm{\hat p}=(\hat p_x,\hat p_y) =-i\hbar \bm{\nabla}$ and Fermi velocity $v_F\approx 10^6$~m$/$s.  
Pauli matrices $\sigma_{x,y}$ 
act in sublattice space, while explicitly written $2\times 2$ matrices refer to Nambu (particle-hole) space throughout.
In particular, $H$ in Eq.~\eqref{dbdg} acts on Nambu spinors
$\Psi({\bm r})= ( u, v)^T$ containing the spin-up electron-like (spin-down hole-like) 
wave function $u$ ($v$) near the $K$ ($K'$) valley, where 
$u$ and $v$ are spinors in sublattice space and ${\bm r}=(x,y)$.
A decoupled identical copy of $H$ with opposite spin is kept implicit  \cite{Beenakker2006}.
The vector potential $\bm{A}=(0,Bx)$ describes a perpendicular homogeneous 
magnetic field $B$ in Landau gauge, where we neglect the typically small Zeeman splitting. 
The potential term in Eq.~\eqref{dbdg} 
also accounts for the chemical potential $\mu$ through the shift $V-\mu\to V$, and
the homogeneous spin-singlet pairing amplitude $\Delta$ (taken real positive below) 
comes from the proximity effect. Note that intrinsic superconductivity in 
graphene \cite{Uchoa2007,Kopnin2008} has not been found experimentally.  
Finally, we neglect Coulomb interactions which
are largely screened off by the proximity-inducing 
superconductor.
In what follows, we measure lengths (wave numbers) in units of the magnetic length 
$l_B$ ($1/l_B$), and energies in units of the cyclotron scale $E_B$, where
\begin{equation}\label{defunits}
l_B=\sqrt{\hbar c/eB}, \quad E_B= \hbar v_F/l_B.
\end{equation}

Equation~\eqref{dbdg} tacitly assumes applied magnetic fields below the critical field of the proximity-inducing superconductor and
 that the Meissner effect is too weak to completely expel the magnetic field from the proximitized graphene layer. 
 In principle, renormalized values of  $B$ and $\Delta$ entering Eq.~\eqref{dbdg} can be obtained from 
self-consistency equations, cf.~Refs.~\cite{Rasolt1992,MacDonald1993}. However,
since coexistence of $B$ and $\Delta$ has already been observed in graphene  \cite{BenShalom2016,Amet2016,Lee2017} and  other 2D 
electron gases \cite{Nichele2017}, we here take them as 
effective parameters and focus on the physics caused by their interplay.

\emph{Chiral representation.---}It is convenient to reformulate Eq.~\eqref{dbdg} using
$4\times 4$ Dirac matrices in the chiral representation, 
$\beta = \left( \begin{array}{cc} 0& -\sigma_0 \\ -\sigma_0 & 0 \end{array}\right)$ and 
$\alpha^j =  \left( \begin{array}{cc}  \sigma_j & 0 \\ 0 & -\sigma_j  \end{array}\right)$,
with $j=1,2,3$ and identity $\sigma_0$ in sublattice space. 
Anticommuting $\gamma^\nu$ matrices are then given by  
$\gamma^0=\beta$ and $\gamma^{j}=\beta \alpha^j$,
where we also define $\gamma^5={\rm diag}(\sigma_0,-\sigma_0)$.
In Landau gauge, Eq.~\eqref{dbdg} is  equivalently expressed as
\begin{equation}\label{bdg}
H= \alpha^1\hat p_x + \alpha^2\left(\hat p_y+x\gamma^5\right)+\gamma^5 V - \beta \Delta.
\end{equation} 
Formally, Eq.~\eqref{bdg} describes 2D Dirac fermions with mass $-\Delta$ 
subject to pseudo-vector and pseudo-scalar potentials: 
the ${\bm A}$ and $V$ terms involve $\gamma^5$. 
Given a BdG eigenstate $\Psi_E=\left( u_E, v_E\right)^T$ with
 energy $E\ge 0$, a particle-hole transformation yields a 
 solution with energy $-E$,
 \begin{equation} \label{phs}
\Psi_{-E}^{}({\bm r})= - \gamma^2 \Psi^{\ast}_E({\bm r})= \left(\begin{array}{c} -\sigma_y v_E^*({\bm r}) 
\\ \sigma_y u_E^*({\bm r}) \end{array}\right).
\end{equation}
Therefore it is sufficient to find solutions with $E\ge 0$, and  
Eq.~\eqref{phs} is a self-conjugation relation for $E=0$. 
For a complete set $(u_\lambda,v_\lambda)^T$ with energies $E_\lambda\ge 0$, the local DOS $\rho(E)$ is defined in a standard way \cite{foot0} and  
can be measured by STM techniques, see Fig.~\ref{fig1}, 
Furthermore, the charge current density ${\bm J}=(J_x,J_y)^T$ corresponding to a given eigenstate is 
\begin{equation}\label{ccur}
{\bm J}_\lambda({\bm r}) = -ev_F \left( u_\lambda^\dagger \bm{\sigma}
u_\lambda^{} + v_\lambda^\dagger \bm{\sigma} v_\lambda^{} \right).
\end{equation}

In what follows, we assume $V=V(x)$ such that  Eq.~\eqref{bdg} enjoys translation invariance along the $y$-direction.  BdG solutions are given by 
$\Psi_{k}({\bm r})=e^{iky}\psi_k(x)$,
where $\psi_k(x)$ is an eigenstate to $H_k$ obtained from $H$ in
Eq.~\eqref{bdg} with $\hat p_y\to k$. 
We now perform a partial (involving only the momentum in $y$-direction) 
Bogoliubov transformation, $\psi_k(x) = M_k \phi_k(x)$, with the 
unitary $4\times 4$ matrix
\begin{eqnarray}\label{mdef}
M_k &=& a_{k,+}- a_{k,-}\gamma^2 =  \left(\begin{array}{cc} a_{k,+} & -\sigma_ya_{k,-}  \\ \sigma_y a_{k,-} & a_{k,+}
\end{array}\right),\\
&& a_{k,\pm } =\nonumber \sqrt{\frac{X_k\pm k}{2X_k}},\quad X_k= \sqrt{k^2+\Delta^2}.
\end{eqnarray}
The  BdG equation, $\tilde H_k\phi_k(x)=E\phi_k(x)$ with
$\tilde H_k= M_k^{-1} H_k M_k$, then involves the transformed Hamiltonian 
\begin{equation}\label{bdg3}
\tilde H_k =\alpha^1\hat p_x + \alpha^2\left(X_k+x\gamma^5\right)+ 
\frac{k+\gamma^2\Delta}{X_k} \gamma^5 V(x).
\end{equation}
For $B=0$ and constant $V$, one has 
plane waves with ${\bm k}=(k_x,k)$ and energy  
 $E_{{\bm k}, \pm}= \sqrt{(\pm\hbar v_F|{\bm k}|+ V)^2+\Delta^2}$~\cite{Beenakker2006}, where the DOS for $E\ge 0$ and $V\ge 0$ is given by 
\begin{equation} \label{tdos1}
\rho(E) =\frac{1}{\pi (\hbar v_F)^2}\times \left\{ 
 \begin{array}{cc} 0, & E<\Delta,\\
 \frac{EV-(E^2-\Delta^2)}{\sqrt{E^2-\Delta^2}} ,& \Delta<E<\sqrt{V^2+\Delta^2},\\
 E-V, & E>\sqrt{V^2+\Delta^2}.
 \end{array} \right. 
\end{equation}
Note that at the Dirac point, i.e., for $V=0$, the usual BCS square-root singularity 
is replaced by a finite jump at $E=\Delta$, with $\rho(E)\sim E$ for $E>\Delta$.

\emph{Exact solution at the Dirac point.---}For $V=0$, we next observe that
$\tilde H_k$ in Eq.~\eqref{bdg3}  coincides with the original Hamiltonian in Eq.~(\ref{bdg}) for $\Delta=0$  and $\hat p_y\to X_k$. As a consequence,  
the entire spectrum coincides with the  
$(k,\Delta)$-independent relativistic Landau energies, $E_{k,n,s}=E_n= \sqrt{2n}E_B$ with 
$n=0,1,2,\ldots$ \cite{CastroNeto2009}.  
On top of the $k$-degeneracy, we have an additional double degeneracy 
with $s=\pm$, see below. 
Eigenstates follow by the above $M_k$ transformation from 
relativistic Landau states. The latter are given by the Nambu spinors
$\phi_{k,n,+}(x)=\left ( {\cal F}_n(x+X_k) , 0\right)^T$ and  
$\phi_{k,n,-} (x)=\left( 0,  \sigma_y {\cal F}_n(x-X_k) \right)^T$,
where  sublattice spinors,    
${\cal F}_{n}(x) = (\frac{1}{\sqrt2})^{1-\delta_{n,0}} \left( {\rm sgn}(n)  \varphi_{|n|-1}, i \varphi_{|n|}\right)^T$, 
are expressed in terms of normalized oscillator eigenfunctions \cite{foot1}.
Note that the usual center-of-mass coordinate $k$ is replaced by $X_k$ 
($-X_k$) for the electron (hole) spinor component, cf.~Eq.~\eqref{bdg3}.
Using Eq.~\eqref{mdef}, eigenstates follow as
\begin{equation}\label{exacteig}
\Psi_{k,n,s=\pm}({\bm r})=
e^{iky} \left( \begin{array}{c} \pm a_{k,\pm} {\cal F}_n(x\pm X_k) \\
a_{k,\mp}\sigma_y{\cal F}_n(x\pm X_k) \end{array}\right). 
\end{equation}
In contrast to the spectrum, these states depend on $\Delta$ and thus
most observables will be sensitive to pairing.  For given $\Psi_{k,n,s}$, 
Eq.~\eqref{phs} yields a mirror state $\Psi_{-k,-n,\pm}({\bm r})=\pm \gamma^2\Psi^*_{k,n,\pm}({\bm r})$ with $E=-E_n$.
For $n=0$, this relation connects $+k$ and $-k$ states, and
one can construct two ($s=\pm$) 1D zero-energy Majorana fields.

\begin{figure}
\centering
\includegraphics[width=0.56\textwidth]{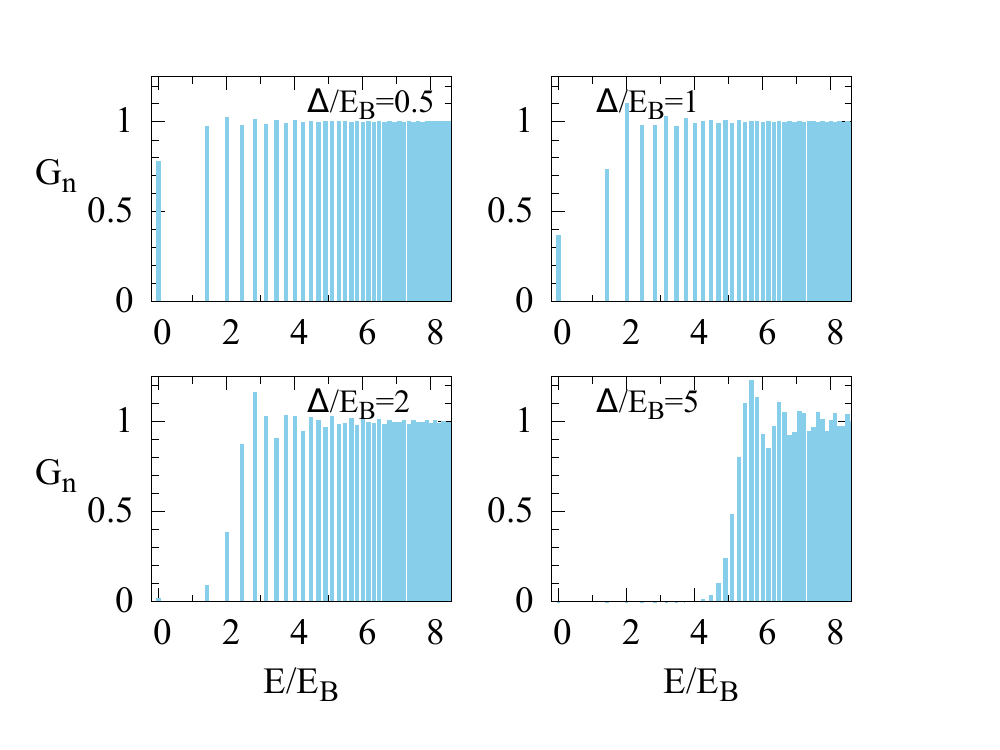} 
\caption{\label{fig2}  
Bar plots of the DOS weights $G_n$ vs Landau energy $E_n$  for different $\Delta/E_B$, see Eqs.~\eqref{tdos2} and \eqref{gdef}.}
\end{figure}  

\emph{Density of states at the Dirac point.---}By using the 
states in Eq.~\eqref{exacteig} and restoring units, we obtain 
an exact integral representation for the DOS  \cite{SM}, 
\begin{eqnarray}\label{tdos2}
 \rho(E)&=& \frac{e^{-(\Delta/E_B)^2}}{\pi l_B^2}\delta(E)+\frac{|E|}{\pi(\hbar 
 v_F)^2}\times \\ \nonumber &\times& 
\int_{-\infty-i0^+}^{+\infty-i0^+} \frac{d\lambda}{2\pi i} 
 e^{i (E^2\lambda-\Delta^2 \tan\lambda )/E_B^2} \cot\lambda,
\end{eqnarray}
which is singular and applies in the distribution sense. 
For $B\to 0$, the asymptotic approximation of
Eq.~\eqref{tdos2} reproduces Eq.~\eqref{tdos1} with $V=0$. 
The bar plots in Fig.~\ref{fig2} show the dimensionless DOS weights
\begin{equation}\label{gdef}
G_n= \pi l_B^2 \int_{E_n-0^+}^{E_n+0^+} dE \rho(E), \quad 
E_n=\sqrt{2n}E_B,
\end{equation} 
characterizing the $\delta(|E|-E_n)$ peaks in the DOS and hence also
the degeneracy per unit area of the energy levels $E_n$.
For $\Delta\to 0$, Eq.~\eqref{tdos2} yields the standard
Landau comb with $G_n=1$.  Figure \ref{fig2} illustrates 
the crossover between the analytically accessible limits 
$\Delta/E_B\to 0$ and $\Delta/E_B\to \infty$,
where low-energy states with $|E|<\Delta$ become gradually depleted 
as $\Delta/E_B$ increases.   The DOS in Fig.~\ref{fig2}
also exhibits oscillatory features in the energy dependence. 

\begin{figure}
\centering
\includegraphics[width=0.53\textwidth]{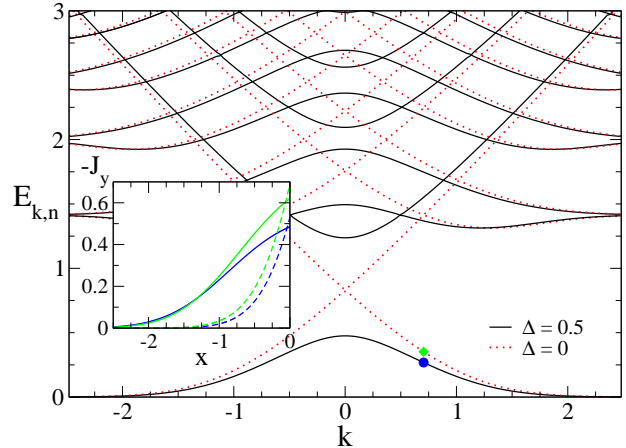} 
\caption{\label{fig3} Edge states for a semi-infinite ($x<0$) graphene sheet 
 with $V=0$ and armchair conditions at $x=0$.  
Main panel: Dispersion relation for $\Delta=0.5E_B$ (solid black) and for $\Delta=0$ (red dotted curves).  Inset:
 Current density $J_y(x)$ [in units of $-ev_F$] vs position $x/l_B$ 
 for the two degenerate eigenstates (solid and dashed curves for $s=+$ and $s=-$, resp.) with 
 $kl_B=0.705$.  Blue (green) curves are for $\Delta/E_B=0.5$ ($\Delta=0$)
 with $E_{k,n,s}/E_B\simeq 0.2683$ ($\simeq 0.3520$), cf.~the
 blue circle (green diamond) in the main panel.   }
\end{figure} 

\emph{Edge states.---}Next we consider a semi-infinite graphene sheet ($x<0$) 
with $V=0$. The boundary is modeled by imposing armchair conditions \cite{Beenakker2008,CastroNeto2009} along the line $x=0$.  
Solutions to the BdG equation are then given in terms of parabolic cylinder
functions $D_p(z)$ \cite{Abramowitz}. The spectrum is obtained by numerically solving
 ${\rm det} [{\bm W}(E)]= 0$, where the matrix ${\bm W}$  
 follows with $\epsilon=E/\sqrt2$, $a_\pm= a_{k,\pm}$ [cf.~Eq.~\eqref{mdef}], and 
 $\tilde D_p^{(\pm)}=D_p\left( \pm \sqrt{2(k^2+\Delta^2)}\right)$
 in the form \cite{SM}
\begin{equation}\label{determinanteq}
\left( \begin{array}{cccc} 
-a_{+}\epsilon\tilde D_{\epsilon^2-1}^{(-)}  & a_+\tilde D_{\epsilon^2}^{(-)} &
 a_-\epsilon \tilde D_{\epsilon^2-1}^{(+)} &a_- \tilde D_{\epsilon^2}^{(+)} \\ 
 a_+ \tilde D_{\epsilon^2}^{(-)}  & -a_+ \epsilon \tilde D_{\epsilon^2-1}^{ (-)}&
 - a_- \tilde D_{\epsilon^2}^{(+)}  & -a_- \epsilon\tilde D_{\epsilon^2-1}^{(+)} \\ 
 a_- \tilde D_{\epsilon^2}^{(-)} & -a_- \epsilon\tilde D_{\epsilon^2-1}^{(-)}&
 a_+ \tilde D_{\epsilon^2}^{(+)} & a_+ \epsilon\tilde D_{\epsilon^2-1}^{(+)} \\
a_-\epsilon \tilde D_{\epsilon^2-1}^{(-)} & -a_-\tilde D_{\epsilon^2}^{(-)} & 
a_+ \epsilon \tilde D_{\epsilon^2-1}^{(+)}& a_+\tilde D_{\epsilon^2}^{(+)}
 \end{array}
  \right)
\end{equation}
The spectrum is shown in Fig.~\ref{fig3}.  For $\Delta=0$, we recover earlier 
results \cite{Brey2006,Abanin2007,Delplace2010} reporting 
chiral edge states.  For $\Delta>0$, electron- and hole-type 
edge states become mixed and the edge state dispersion exhibits gaps near $k=0$.  
Turning to the current density \eqref{ccur}, the current flows along the $y$-direction only, $J_x=0$. The  respective profile, $J_y(x)$, is illustrated for the
 two degenerate states with $k=0.705$ and lowest energy in the inset of 
 Fig.~\ref{fig3}.  Since the current density has a pronounced peak near $x=0$ 
 and a specific sign, we have unidirectional edge states also for $\Delta>0$.
 However, the overall current becomes smaller with increasing
 $\Delta$, cf.~Fig.~\ref{fig3}.

\begin{figure}
\centering
\includegraphics[width=0.53\textwidth]{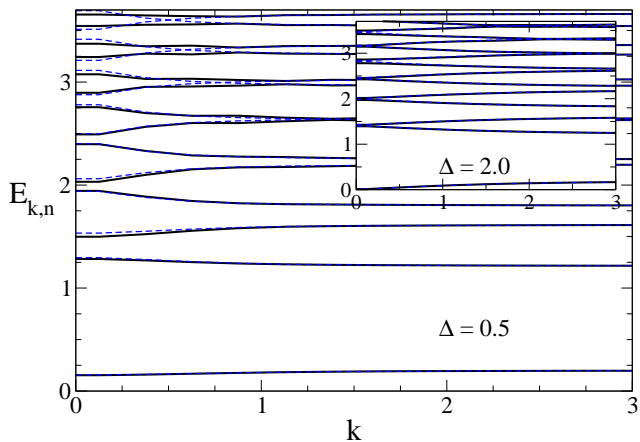} 
\caption{\label{fig4} Dispersion relation for an infinite graphene sheet with
potential $V=0.2E_B$ for $\Delta=0.5E_B$ (main panel) and $\Delta=2E_B$ (inset).  
Since $E_{-k,n}=E_{k,n}$, only $k\ge 0$ is shown.
Solid black and dashed blue curves refer to numerical diagonalization  and 
perturbative results [Eq.~\eqref{correction}], respectively.}
\end{figure}

\emph{Going away from the Dirac point.---}Let us briefly address the case $V\ne 0$, where
numerical  diagonalization of the BdG equation using Landau states as 
basis shows that a (chemical) potential shift causes dispersion, see Fig.~\ref{fig4}.     
Notably, most features in Fig.~\ref{fig4} can be understood by expanding 
around the $V=0$ solution \eqref{exacteig} using
 the term $\sim V$ in Eq.~\eqref{bdg3} as small perturbation. 
Writing $E_{k,n,s}=E_n+\delta E_{k,n,s}$, first-order degenerate 
perturbation theory yields the correction 
\begin{equation}\label{correction}
\delta E_{k,n,\pm}= \pm\frac{ |V| }{X_k}
 \sqrt{k^2+\Delta^2S^2_{k,n}},
\end{equation}
where the overlap between Landau states ${\cal F}_n$ centered at 
$+ X_k$ and $-X_k$ is encoded by  $S_{k,n}$.  Explicitly, we find
$S_{k,0}=e^{-X_k^2}$ and $S_{k,n>0}=
\frac12 e^{-X_k^2} [L_{n-1}(2X_k^2)+L_n(2X_k^2)]$
with the Laguerre polynomials $L_n$ \cite{Abramowitz}.
For $|k|\gg \Delta$, Eq.~\eqref{correction} yields a uniform shift $\pm |V|$ of all Landau energies, while for $k=0$, the correction simplifies to $\pm |V S_{0,n}|$, where 
$S_{0,n}$ oscillates when changing $n$. 

\emph{Crossed electric and magnetic fields.---}We finally also
include an in-plane electric field ${\cal E}$ by putting
$V=e{\cal E} x$. With the dimensionless parameter 
$\varepsilon= (c/v_F) {\cal E}/B$, we consider the regime $|\varepsilon|<1$.  
The corresponding $\Delta=0$ problem has been solved analytically  
by a Lorentz boost into the reference frame with vanishing electric field (${\cal E}'=0)$
\cite{Lukose2007}.  Remarkably, such a strategy also admits an exact 
solution for  $\Delta\ne 0$:  First, we write down the spinor 
transformation law, $\psi=S\psi'$ with 
$S= \cosh(\eta/2) - \sinh(\eta/2) \gamma^0\gamma^2$, where the Lorentz angle
$\eta=\tanh^{-1}\varepsilon$ defines the frame with ${\cal E}'=0$.
Next, using the parameter $\zeta\equiv (1-\varepsilon^2)^{1/4}$, we rescale (i) the $x$-coordinate,
$x'=\zeta x$, (ii) the wave number,
$k'= (k+\varepsilon E)/\zeta^3$,  (iii) energy,  
 $E'= (E+\varepsilon k)/\zeta^3$, and (iv) the proximity gap,
$\Delta'=\Delta/\zeta$.  With these rescalings and  $X_k'=\sqrt{ k^{\prime 2}+ \Delta^{\prime 2}}$, cf.~Eq.~\eqref{mdef},  
the BdG equation in the new frame coincides with the $V=0$ problem solved above.  
 Transforming the solution, Eq.~\eqref{exacteig}, back to the lab frame 
 and restoring units, we obtain the $\Delta$-independent spectrum
 \begin{equation}\label{snakestates}
 E_{k,n,s}= -\hbar \varepsilon v_F k + {\rm sgn}(n)\sqrt{2|n|}\zeta^{3} E_B,
 \end{equation} 
 where $n$ runs over all integers and $k$ is restricted to those values with $E_{k,n,s}\ge 0$.  Each level is two-fold degenerate ($s=\pm$), and the 
 corresponding eigenstates are  
\begin{eqnarray} 
\Psi_{k,n,\pm}({\bm r}) &=& e^{iky}\zeta^{3/2} 
\Biggl [ \cosh(\eta/2)\left( \begin{array}{c} \pm a_{k',\pm} 
{\cal F}_n(x'\pm X_k')\\ a_{k',\mp}\sigma_y {\cal F}_n (x'\pm X'_k)
\end{array} \right)   \nonumber \\ 
\label{snakest}
&+& \sinh(\eta/2) \left( \begin{array}{c} \mp a_{k',\pm}\sigma_y {\cal F}_n(x'\pm X'_k)  
\\ a_{k',\mp} {\cal F}_n(x'\pm X'_k)  \end{array}\right) \Biggr ],
\end{eqnarray}
States with negative energy follow from Eq.~\eqref{phs}, and  
for $\varepsilon=0$, Eq.~\eqref{snakest} reduces to Eq.~\eqref{exacteig}.  

In the normal ($\Delta=0$) case, 
so-called snake states exist near the interface between $V>0$ and $V<0$ regions \cite{Liu2015,Cohnitz2016,Tatch2015,Rickshaus2015}
which are semiclassically described by snake-like orbits propagating 
along the interface (here the $y$-direction) with  
velocity $c{\cal E}/B=\varepsilon v_F$.  In the superconducting case ($\Delta>0$),
the spectrum in Eq.~\eqref{snakestates} suggests that unidirectional 
snake states remain well defined and propagate with the same 
snake velocity as for $\Delta=0$. In particular,
for $n=0$, these states are localized near the line $x=0$. 
Computing the total charge current carried by a given state along the $y$-direction,
$I=\int dx J_y(x)$, Eqs.~\eqref{ccur} and \eqref{snakest} yield the analytical
result $I(\Delta)/I(0)=1/\sqrt{1+(\Delta'/k')^2}$.  Similar to the above edge state case,
we thus find that the magnitude of the current becomes gradually suppressed 
with increasing $\Delta$. 

\emph{Conclusions.---}We have studied electronic properties of graphene 
monolayers in an orbital magnetic field when also proximity-induced pairing
correlations are present.  Remarkably, at the Dirac point, the energy spectrum 
is independent of $\Delta$, but observables may still show 
pronounced pairing effects since eigenstates depend on $\Delta$.  
We hope that our work will stimulate experimental and further theoretical work
on the coexistence of magnetism and superconductivity in graphene. 
  
\acknowledgments
We thank T. Kontos and C. Sch\"onenberger for helpful discussions and
acknowledge support by the DFG network CRC TR 183 (project C04).

\newpage

\appendix

\section{Density of states at Dirac point}

We first discuss the derivation of Eq.~(10) in the main text. Below we set $a=\Delta/E_B$. 
Using the exact $V=0$ states in Eq.~(9), the local DOS takes the form 
\begin{eqnarray}\nonumber
\rho(E)&=&2 g_{0}\delta(E)+ \sum_{n>0}
(g_{n-1}+g_{n})  \delta\left(|E|-\sqrt{2n}E_B\right),\\
\label{gn}
g_{n} &=& \int \frac{dk}{2\pi} \varphi^2_n \left(\sqrt{(kl_B)^2+a^2}\right)
=\frac{1}{2\pi l_B^2}I_n(a),\\ \nonumber
I_n(a) &=&\frac{1}{\sqrt{\pi}2^n n!} 
\int_{a^2}^\infty \frac{du}{\sqrt{u-a^2}} H^2_n\left(\sqrt{u}\right) e^{-u}.
\end{eqnarray}
For $\Delta=0$, we have $I_n(0)=1$ and the Landau comb is reproduced.  Moreover, $I_0(a)$ yields the $\delta(E)$ prefactor in Eq.~(10).  
We thus focus on the local DOS for $|E|>0$.  With  $D$ denoting 
an effective high-energy bandwidth, where eventually the 
limit $D\to \infty$ has to be taken, we can rewrite Eq.~(\ref{gn}) as
\begin{eqnarray}\nonumber  
&& \rho(E)  = \frac{1}{2\pi l^2_B} 
\sum_{n>0} e^{-2n(E_B/D)^2}
\left[ I_{n-1}(a) + I_{n}(a) \right] \times \\ && \qquad \qquad \qquad \times\  \delta\left(|E| - \sqrt{2n}E_B\right)\\
&&= \nonumber \frac{|E|}{\pi l^2_B} \sum_{n>0} ( I_{n-1} + I_{n} ) e^{-2n(E_B/D)^2} 
\int_{-\infty}^{+\infty} \frac{d\lambda}{2\pi}
 e^{i\lambda(E^2-2nE^2_B) }
\end{eqnarray}
with $\delta(E^2-2n E_B^2)= (2|E|)^{-1}\delta(  |E|-\sqrt{2n}E_B)$ and 
an integral representation of the $\delta$-function. 
Exchanging sum and integral, measuring $E$ in units of $\Delta$ and rescaling 
$\lambda \to \lambda/\Delta^2$, we find 
\begin{eqnarray}\label{almostfinal}
\rho (E) &=& \frac{\Delta |E|}{\pi (\hbar v_F)^2} e^{-(E/\tilde D)^2} \times
\\ &\times& \int_{-\infty}^{+\infty}
\frac{d\lambda}{2\pi a^2}  e^{i\tilde\lambda E^2} 
\nonumber \left( e^{-2i\tilde\lambda} +1 \right) {\cal G}_a(\tilde\lambda),
\end{eqnarray}
where we define $\tilde D=D/\Delta$,
\[
{\cal G}_a(\tilde\lambda)=\sum_{n\geq 0} I_{n}(a) e^{-2i\tilde\lambda  n }, \quad
\quad \tilde \lambda=\frac{1}{a^2}\left(\lambda-\frac{i}{\tilde D^2}\right).
\]
Next, using $\left|e^{-2i\tilde\lambda} \right|<1$ and the Poisson kernel \cite{Abramowitz}, 
we sum up the series,
\begin{eqnarray}\nonumber
{\cal G}_a(\tilde\lambda) &=& \sum_{n \geq 0} \frac{1}{\sqrt{\pi} } \int_{a^2}^\infty 
\frac{du}{\sqrt{u-a^2}} \frac{H^2_n(\sqrt{u})}{2^nn!} e^{-u}
e^{-2i\tilde\lambda n}   \\
&=&\frac{1}{\sqrt{\pi}} \int_{a^2}^\infty  \frac{du}{\sqrt{u-a^2}} \nonumber
\frac{\exp\left(u \frac{2e^{-2i\tilde\lambda}}{1+e^{-2i\tilde\lambda}}\right)}
{(1-e^{-4i\tilde\lambda})^{1/2}}e^{-u} \\ \label{auxil}
&=&  \frac{1}{1-e^{-2i\tilde\lambda}}
\exp \left(- \frac{1-e^{-2i\tilde\lambda}}{1+e^{-2i\tilde\lambda}}a^2\right).
\end{eqnarray}
 Inserting Eq.~(\ref{auxil}) into Eq.~(\ref{almostfinal}), we obtain
 \begin{eqnarray}\nonumber
&& \rho (E) = \frac{\Delta |E|}{\pi (\hbar v_F)^2} e^{-(E/\tilde D)^2} \times
\\ && \quad \times \int_{-\infty-i/\tilde D^2}^{+\infty-i/\tilde D^2} 
\frac{d\lambda}{2\pi i} \frac{ e^{i\lambda E^2-a^2 \tan(\lambda/a^2)} }{a^2\tan(\lambda/a^2)}.
 \end{eqnarray}
 Restoring units, letting $D\to \infty$, and including the $E=0$ peak, 
 we arrive at Eq.~(10) in the main text.  
 
\section{On the determinantal condition}

We here consider the semi-infinite case ($x<0$) with $V=0$ and armchair 
boundary conditions imposed on the line $x=0$. For given wave number $k$ 
and energy $E$, using the parabolic cylinder functions $D_p(z)$ \cite{Abramowitz},
general solutions of the BdG equation that are normalizable for $x<0$ are given by
the Nambu spinors
\begin{eqnarray}\label{appb1}
\Psi_{k,E} ({\bm r}) &=& c_1 e^{iky}\left( \begin{array}{c}a_{+}{\cal F}_{X_k,E}(x) \\
a_{-} \sigma_y{\cal F}_{X_k,E}(x)\end{array} \right) + \\ \nonumber & +&
c_2 e^{iky}\left( \begin{array}{c}-a_{-}{\cal F}_{-X_k,E}(x) \\
a_{+} \sigma_y{\cal F}_{-X_k,E}(x)\end{array} \right),
\end{eqnarray}
with complex coefficients $c_{1,2}$, the numbers $a_{\pm}\equiv a_{k,\pm}$
in Eq.~(6), and the sublattice spinors  ($p\equiv E^2/2$)
\begin{equation}
{\cal F}_{\pm X_k,E}(x)= \left( \begin{array}{c} 
-\frac{E}{\sqrt{2}} D_{p-1}\left(-\sqrt2 (x\pm X_k)\right) \\
i D_{p}\left(-\sqrt2 (x\pm X_k)\right) \end{array} \right).
\end{equation}
We now impose armchair boundary conditions at $x=0$,
\begin{equation}\label{abc}
\psi_A(0,y)+\psi'_A(0,y) =0,\quad \psi_B(0,y)+\psi'_B(0,y) =0,
\end{equation}
where the sublattice spinor components 
$\psi_{A/B}({\bm r}) \ [\psi'_{A/B}({\bm r})]$ characterize an 
electron at the $K$ [$K'$] valley and Eq.~(\ref{abc}) has to be satisfied for all $y$.  
Next we note that the upper Nambu spinor component in Eq.~(\ref{appb1}) contains  
$\psi_{A/B}({\bm r})$  for an electron at the $K$ valley 
with wave number $k$ and energy $E$, while the lower component 
of Eq.~(\ref{appb1}) contains the complex conjugate of 
$\psi'_{A/B}({\bm r})$ for an electron at the $K'$ valley 
with wave vector $-k$ and energy $-E$.
In order to satisfy Eq.~(\ref{abc}), we thus have to consider superpositions 
of $\pm k$ states with the same energy $E$.  
Using complex coefficients $d^*_{1,2}$ to parametrize the partner states 
with wave number $-k$ and the same energy $E$, see Eq.~(\ref{appb1}), 
and using ${\cal F}^*_{\pm X_k,-E}=-{\cal F}_{\pm X_k,E}$,  Eq.~(\ref{abc})
yields the relations
\begin{eqnarray}\label{ffff}
&& c_1 a_+{\cal F}_{X_k,E} -c_2 a_- {\cal F}_{-X_k,E} \\ \nonumber
&& + \ d_1 a_+\sigma_y {\cal F}_{X_k,E}  + d_2 a_- \sigma_y{\cal F}_{-X_k,E} =0, \\
\nonumber
&& c_1 a_-\sigma_y {\cal F}_{X_k,-E} +  c_2 a_+ \sigma_y{\cal F}_{-X_k,-E} \\
\nonumber
&&  -  \ d_1 a_-{\cal F}_{X_k,-E}  + d_2 a_+ {\cal F}_{-X_k,-E}  =0,
\end{eqnarray}
where all sublattice spinors ${\cal F}$ are taken at $x=0$.  The relations (\ref{ffff}) 
result in four equations for the four variables ($c_1,d_1,c_2,d_2$). 
We thus arrive at the matrix ${\bm W}(E)$ in Eq.~(12).
For $\Delta=0$, the corresponding determinantal condition simplifies to 
$p D_{p-1}^2(-\sqrt2 k) = D_p^2(-\sqrt2 k)$.


\begin{thebibliography}{99}
\bibitem{foot0}
Using $f_\lambda=2-\delta_{E_\lambda,0}$, the local DOS is given by
$\rho(E)=\sum_\lambda f_\lambda \left[ |u_\lambda({\bm r})|^2 \delta(E-E_\lambda) + 
|v_\lambda({\bm r})|^2\delta(E+E_\lambda)\right]$ and expected to be 
independent of ${\bm r}$ for a homogeneous system.
For simplicity, we evaluate it for $x=y=0$ here.
\bibitem{SM}
See the appendices, where we
provide details about the derivation of Eqs.~\eqref{tdos2} and \eqref{determinanteq}.
\end{thebibliography}
\end{document}